\documentclass[conference,10pt]{IEEEtran}

\IEEEoverridecommandlockouts
\def\BibTeX{{\rm B\kern-.05em{\sc i\kern-.025em b}\kern-.08em
    T\kern-.1667em\lower.7ex\hbox{E}\kern-.125emX}}

\usepackage[utf8]{inputenc}
\usepackage{graphicx}
\usepackage{color}
\usepackage{xcolor}
\usepackage{cite}
\usepackage{amsmath}
\usepackage[caption=false]{subfig}
\usepackage{amssymb}
\usepackage{comment}
\usepackage{mathtools}
\usepackage{tabulary}
\usepackage{acronym}
\usepackage{upgreek}
\usepackage{algorithm}
\usepackage{algpseudocode}
\usepackage{placeins}
\usepackage[super]{nth}
\usepackage{dsfont}
\usepackage{amsthm}
\usepackage{array}

\newtheorem{corollary}{Corollary}
\newtheorem{theorem}{\bf Theorem}

\newtheorem{remark}{Remark}

\newcommand{\myvec}[1]{\boldsymbol{#1}}

\begin{document}
\title{Asymptotically Optimal Closed-Form Phase Configuration of $1$-bit RISs via Sign Alignment}



\author{
    Kyriakos Stylianopoulos$^{1}$,
    Panagiotis Gavriilidis$^{1}$,
    and George C. Alexandropoulos$^{1,2}$ \\
    \small
    $^1$ Department of Informatics and Telecommunications, National and Kapodistrian University of Athens, Greece \\
    $^2$ Department of Electrical and Computer Engineering, University of Illinois Chicago, IL, USA \\
    {\{kstylianop, pangavr, alexandg\}@di.uoa.gr}
    \thanks{This work has been supported by the SNS JU project TERRAMETA under the EU's Horizon Europe research and innovation programme under Grant Agreement No 101097101, including top-up funding by UKRI under the UK government's Horizon Europe funding guarantee.}
    \vspace{-0.2cm}
}

\maketitle

\begin{abstract}
While Reconfigurable Intelligent Surfaces (RISs) constitute one of the most prominent enablers for the upcoming sixth Generation (6G) of wireless networks, the design of efficient RIS phase profiles remains a notorious challenge when large numbers of phase-quantized unit cells are involved, typically of a single bit, as implemented by a vast majority of existing metasurface prototypes. In this paper, we focus on the RIS phase configuration problem for the exemplary case of the Signal-to-Noise Ratio (SNR) maximization for an RIS-enabled single-input single-output system where the metasurface tunable elements admit a phase difference of $\pi$ radians. We present a novel closed-form configuration which serves as a lower bound guaranteeing at least half the SNR of the ideal continuous (upper bound) SNR gain, and whose mean performance is shown to be asymptotically optimal. The proposed sign alignment configuration can be further used as initialization to standard discrete optimization algorithms. A discussion on the reduced complexity hardware benefits via the presented configuration is also included. Our numerical results demonstrate the efficacy of the proposed RIS sign alignment scheme over iterative approaches as well as the commonplace continuous phase quantization treatment.
\end{abstract}

\vspace{-0.3cm}
\section{Introduction}

The technology of Reconfigurable Intelligent Surfaces (RISs) is posed as one of the key enablers for next sixth Generation (6G) of wireless networks due to its dynamic connectivity enhancements with minimal deployment and operation costs~\cite{ABoI_EURASIP,RISsurvey2023}. In fact, theoretical results have shown that an RIS of $N$ unit cells can offer an $N^2$-fold increase in the end-to-end channel magnitude by optimally configuring the phase shifts of its comprising elements \cite{PLP21}. Such benefits are further amplified by the metasurfaces' near-passive energy consumption and minute manufacturing cost per unit element. 
In an effort to reduce hardware complexity, however, a vast majority of developed prototypes has been designed with elements admitting discrete phase shift values, often with a discretization of a single bit.
Not only are the optimal gains of finite discrete RIS configurations subpar to the theoretical $N^2$ gain (e.g., \cite{WZ20, SA22, LHG23}), but also this design principle casts the optimization problem of the surface's phase shifts into the area of discrete optimization~\cite{Risdiscrete2018,Beams_Hw}. In fact, most problem formulations associated with RIS phase configuration are NP-hard and the approximate optimization procedures employed often lead to degraded performance, while requiring impractical computational costs~\cite{PLP21, YCY23}, considering the fact that the number of elements of RISs is expected to be in the order of thousands, notwithstanding their lack of theoretical guarantees.

Motivated by the above, this work specifically tackles the problem of maximization of the Signal-to-Noise Ratio (SNR) for an RIS-enabled Single Input Single Output (SISO) system, when each element of the RIS admits $1$-bit quantized phase shifts with a $\pi$ phase difference between them.
By re-expressing the SNR maximization problem in this particular case, we derive a closed-form configuration vector for the RIS elements that is guaranteed to always achieve more than half of the achievable gain, while its expected gain is proved to be at least $N^2/4$ when Rayleigh channel fading or Line of Sight (LoS) conditions are encountered. We furthermore proceed to use this configuration as initial point to a standard discrete optimization approach offering both performance improvements and faster convergence.
Our numerical evaluation demonstrates that the proposed 
closed-form approach outperforms the standard baseline of discretizing the optimal continuous RIS configuration, as commonly used.

\textbf{Notation}: 
Matrices (vectors) are expressed in bold uppercase (bold lowercase) typeface.
${\rm sign}(\cdot)$ returns the sign of its argument in $\{-1, +1\}$.
Element-wise multiplication and expectation are denoted by $\odot$ and $\mathbb{E}[\cdot]$, respectively.
The real (complex) normal distribution is expressed by $\mathcal{N}$ ($\mathcal{CN}$) and the central Chi-squared distribution with \(k\) degrees of freedom is denoted as \(\mathcal{\chi}^2_{k}\).
$\mathfrak{Re}(\cdot)$ and $\mathfrak{Im}(\cdot)$ return the real and imaginary parts of a complex quantity, $\angle$ denotes the phase of a complex number in Euler form, while $\jmath$ denotes the imaginary unit. The transpose, conjugate, and Hermitian transpose of $\mathbf{A}$, are denoted as \(\mathbf{A}^{\rm T}\), \(\mathbf{A}^{\ast}\), and \(\mathbf{A}^{\rm H}\), respectively.
Finally, lowercase subscripts denote elements of vectors or matrices.

\section{System Model and Design Objective}
Consider a SISO system in the presence of an RIS with $N$ unit cells, where the channel responses for the Transmitter (TX) $\to$ RIS and the RIS $\to$ Receiver (RX) links are denoted by $\boldsymbol{h} \in \mathbb{C}^{N \times 1}$ and $\boldsymbol{g}^{\rm H} \in \mathbb{C}^{1 \times N}$, respectively.
For simplification of presentation, and to present theoretical results in line with the relevant literature, assume there is no direct TX $\to$ RX link.
Let the diagonal matrix $\boldsymbol{\Phi} \in \mathbb{C}^{N \times N}$ include the RIS phase profile, so that each $n$-th unit cell ($n=1,2,\ldots,N$) with configuration $\boldsymbol{\theta}_n$ is modeled as $\boldsymbol{\Phi}_{n,n} = \exp(\jmath \boldsymbol{\theta}_n)$.
By further denoting as $\rho$ the value of the transmit SNR and as \({\rm PL}\) the multiplicative path loss of the two links, then the SISO SNR optimization objective can be formulated as follows~\cite{WZ20}:
\begin{align}
    \mathcal{OP}_1:&~ \underset{\boldsymbol{\theta}} {\max}\left\{ \rho {\rm PL}|\boldsymbol{g}^{\rm H} \boldsymbol{\Phi} \boldsymbol{h}|^2\right\} = \underset{\boldsymbol{\theta}} {\max} \left\{ {\rho {\rm PL}|(\boldsymbol{h} \odot \boldsymbol{g}^{\ast})^{\rm T}  \boldsymbol{\phi}|^2} \right\}\nonumber \\ \nonumber
    &\hspace{0.16cm}{\rm s.t.}~~ \boldsymbol{\theta}_n \in \mathcal{F},~n=1,2,\ldots,N, 
\end{align}
where $\boldsymbol{\phi}$ is the $N$-element vectorized form of the main diagonal of $\boldsymbol{\Phi}$, and \(\boldsymbol{h}\) and \(\boldsymbol{g}\) are normalized such that \(\mathbb{E}[\boldsymbol{h}^{\rm H}\boldsymbol{h}] = \mathbb{E}[\boldsymbol{g}^{\rm H}\boldsymbol{g}] = N\). In addition, $\mathcal{F}$ represents the set of available quantized phase shift values for each RIS element.


In the remainder of the paper, we will be assuming the availability of Channel State Information (CSI) and, for ease of notation, that $\mathcal{F} = \{0, \pi \}$, so that the phase shift for each RIS element aligns with the axis of the real numbers of the complex plane. To this end, $\phi^0\triangleq\exp(\jmath 0) = 1$ and $\phi^{\pi}\triangleq\exp(\jmath \pi) = -1$, therefore, $\boldsymbol{\phi} \in \{-1, +1\}^N$. 
It is noted that the derivations hold for any two phase shift values with a difference of $\pi$ radians, through a rotation of axes.
Based on the above and by denoting the concatenated channel vector as $\boldsymbol{c} \triangleq (\boldsymbol{h} \odot \boldsymbol{g}^{\ast}) \in \mathbb{C}^{N \times 1}$,  $\mathcal{OP}_1$ can be re-formulated as follows:
\begin{align}
    \mathcal{OP}_2:&~ \underset{\boldsymbol{\phi}}\max\, \gamma(\boldsymbol{\phi}) \triangleq |\boldsymbol{c}^{\rm T} \boldsymbol{\phi} |^2  \nonumber \\
    &~ \hspace{1.1cm}=\underbrace{\left( \sum_{i=1}^{N} \boldsymbol{\phi}_i\mathfrak{Re}(\boldsymbol{c}_i) \right)^2}_{\triangleq A(\boldsymbol{\phi})}+ \underbrace{\left( \sum_{i=1}^{N} \boldsymbol{\phi}_i \mathfrak{Im}(\boldsymbol{c}_i) \right)^2}_{\triangleq B(\boldsymbol{\phi})} \nonumber\\
    &\hspace{0.16cm}{\rm s.t.}~~\boldsymbol{\phi} \in \{-1, +1\}^N, \nonumber 
\end{align}
where $\rho$ and \({\rm PL}\) have been dropped without affecting the equivalence between the $\mathcal{OP}_1$ and $\mathcal{OP}_2$ problems.

\section{Closed-Form RIS Phase Configuration}
We commence with the following remark elaborating on the optimization of the two terms appearing in $\mathcal{OP}_2$'s objective.
\begin{remark}\label{remark:two-sums-max}
The maximum values of functions $A(\boldsymbol{\phi})$ and $B(\boldsymbol{\phi})$ in $\mathcal{OP}_2$ for the feasible values of $\boldsymbol{\phi}$ appearing in its constraint are given respectively by $A_{\max} \triangleq (\sum_{i=1}^{N} |\mathfrak{Re}(\boldsymbol{c}_i)|)^2$ and $B_{\max} \triangleq (\sum_{i=1}^{N} |\mathfrak{Im}(\boldsymbol{c}_i)|)^2$, which would be respectively achieved via the RIS phase configurations
$\boldsymbol{\phi}^{\text{Re}} \triangleq {\rm sign}(\mathfrak{Re}(\boldsymbol{c}))$ and $\boldsymbol{\phi}^{\text{Im}} \triangleq {\rm sign}(\mathfrak{Im}(\boldsymbol{c}))$.
\end{remark}
\vspace{-0.1cm}

In Algorithm~\ref{alg:1}, we present a simple approach that selects between the two configurations $\boldsymbol{\phi}^{\text{Re}}$ and $\boldsymbol{\phi}^{\text{Im}}$ the one that maximizes the dual-summand objective of $\mathcal{OP}_2$, i.e., the received SNR. The computational time of this approach is $\Theta(N)$, which is optimal. In the sequel, we provide a theoretical analysis of this Sign Alignment (SA) scheme for RIS phase configuration along with guarantees on its performance. 

\vspace{-0.15cm}
\subsection{Theoretical Analysis}\label{sec:sign-alignment-analysis}
\begin{theorem}[Lower bound for the achievable instantaneous SNR]\label{theorem:deterministic-LB}
When the RIS phase configuration is set to $\boldsymbol{\phi}^{\text{SA}}$ via Algorithm~\ref{alg:1}, the achievable instantaneous SNR $\gamma(\boldsymbol{\phi}^{\text{SA}})$ is lower bounded by the quantity:
\vspace{-0.1cm}
\small{
\begin{equation} \label{eq:lb1}
    f_{\rm LB}(\boldsymbol{\phi}^{\text{SA}}) \!\triangleq\! \max\! \big\{\!\big( \sum\nolimits_{i=1}^{N} |\mathfrak{Re}(\boldsymbol{c}_i)| \big)^{\!\!2}, \big( \sum\nolimits_{i=1}^{N} |\mathfrak{Im}(\boldsymbol{c}_i)| \big)^{\!\!2}\!\big\}\!.
\end{equation}
}
\begin{proof}
    It follows directly from the selection criterion in Step~$3$ of Algorithm~\ref{alg:1} that either $A(\boldsymbol{\phi})$ or $B(\boldsymbol{\phi})$ are maximized, and the maximum values of $A(\boldsymbol{\phi})$ and $B(\boldsymbol{\phi})$ are obtained using the results in Remark~\ref{remark:two-sums-max}.
\end{proof}
\end{theorem}
\begin{algorithm}[t]
\caption{RIS Phase Configuration via Sign Alignment}\label{alg:1}
\begin{algorithmic}[1]
\Require The CSI vector $\boldsymbol{c}$.
\State Set $\myvec{\phi}^{\text{Re}} \gets \text{sign}(\mathfrak{Re}(\boldsymbol{c}))$.
\State Set $\myvec{\phi}^{\text{Im}} \gets \text{sign}(\mathfrak{Im}(\boldsymbol{c}))$.
\State \textbf{Select} $\boldsymbol{\phi}^{\text{SA}} = {\rm arg}\max_{\boldsymbol{\phi} \in \{\myvec{\phi}^{\text{Re}}, \boldsymbol{\phi}^{\text{Im}}\}} \gamma(\boldsymbol{\phi})$.
\end{algorithmic}
\end{algorithm}

\begin{corollary}[Lower bound for the achievable SNR w.r.t. optimal configuration]
Let $\boldsymbol{\phi}^{\star}$ be the RIS phase configuration obtained by optimally solving $\mathcal{OP}_2$, then $\gamma(\boldsymbol{\phi}^{\text{SA}}) \geq 0.5\gamma(\boldsymbol{\phi}^{\star})$.
\begin{proof}
It follows from $\mathcal{OP}_2$'s dual-summand objective, Theorem~\ref{theorem:deterministic-LB}, and the fact that $A(\boldsymbol{\phi}),B(\boldsymbol{\phi})>0$.
\end{proof}
\end{corollary}

\begin{theorem}[Lower bound for the expected achievable SNR under Rayleigh fading]\label{theorem:stochastic-LB}
When both the links TX $\to$ RIS and RIS $\to$ RX experience Rayleigh fading conditions, i.e., $\boldsymbol{h}_i,\boldsymbol{g}_i\overset{i.i.d.}{\sim} \mathcal{CN}(0, 1)$ $\forall$$n=1,2,\ldots,N$, the expectation of the achievable SNR, $ \mathbb{E}[\gamma(\boldsymbol{\phi}^{\text{SA}})]$, is lower bounded by the quantity:
\begin{equation} \label{eq:expected-LB-Rayleigh}
    \bar{f}_{\rm LB}(\boldsymbol{\phi}^{\text{SA}}) \triangleq 0.25N^2.
\end{equation}    
\begin{proof}
Without loss of generality, let us assume that $( \sum_{i=1}^{N} |\mathfrak{Re}(\boldsymbol{c}_i)| )^2 > ( \sum_{i=1}^{N} |\mathfrak{Im}(\boldsymbol{c}_i)| )^2$. Then, following Step~$3$ of Algorithm~\ref{alg:1} results in $\boldsymbol{\phi}^{\text{SA}} = \myvec{\phi}^{\text{Re}}$, yielding:
\begin{align}
   & \mathbb{E} \left[ \gamma(\boldsymbol{\phi}^{\text{SA}}) \right] 
    = \mathbb{E} \left[ A(\myvec{\phi}^{\text{Re}}) + B(\myvec{\phi}^{\text{Re}}) \right] 
   \geq \mathbb{E} \left[ A(\myvec{\phi}^{\text{Re}}) \right] \\
   & = \mathbb{E} \left[ A_{\max} \right] 
    = \mathbb{E} \left[ \left( \sum_{i=1}^{N} |\mathfrak{Re}(\boldsymbol{c}_i)| \right)^2  \right] \\
    & \geq \mathbb{E}^2 \left[ \sum_{i=1}^{N} |\mathfrak{Re}(\boldsymbol{c}_i)| \right] 
    =  \sum_{i=1}^{N} \mathbb{E}^2 \left[ |\mathfrak{Re}(\boldsymbol{c}_i)| \right] . \label{eq:Rayleigh-LB-last-step}
\end{align}
We next express $\boldsymbol{h}_i = x_i + \jmath y_i$ and $\boldsymbol{g}_i^{\ast} = z_i + \jmath w_i$ with \(x_i,\,y_i,\,z_i,w_i\overset{i.i.d.}{\sim} \mathcal{N}(0,0.5)\), which leads to the formulation:
\begin{align}\label{eq:expectation-proposition}
    \mathbb{E} \left[ |\mathfrak{Re}(\boldsymbol{c}_i)| \right] = \mathbb{E} \left[ |\mathfrak{Re}(\boldsymbol{h}_i \boldsymbol{g}^{*}_i)| \right]= \mathbb{E} \left[ |x_i z_i - y_i w_i| \right].
\end{align}

It is well known that the product of the two normally distributed Random Variables (RVs) \(x_i\) and \(z_i\) follows the distribution of the difference of two chi-squared RVs, since \(x_iz_i=0.25\left(\left(x_i+z_i\right)^2 - \left(x_i-z_i\right)^2\right)\). Furthermore, since both terms inside the parentheses have the same variance, it holds that they can be thought of as i.i.d. \(\mathcal{\chi}^{2}_1\) RVs, i.e., their Pearson correlation coefficient in \cite[Sec.~4.B]{book_for_gaussian_distributions} is zero.
By introducing the terms \(Q_i \triangleq  (x_i + z_i)^2 + (y_i - w_i)^2  \) and \(R_i  \triangleq (x_i - z_i)^2 + (y_i + w_i)^2\), \eqref{eq:expectation-proposition} can be re-expressed as:
\begin{equation}\label{eq:Chi-squared reform}
\mathbb{E}\left[ |x_i z_i - y_i w_i| \right] = 0.25 \mathbb{E}\left[ |Q_i - R_i| \right],
\end{equation}
where \(Q_i\) and \(R_i\) are i.i.d. \(\mathcal{\chi}^{2}_2\) RVs.
Then, capitalizing on the results in~\cite[Sec.~4.C]{book_for_gaussian_distributions}, the probability density function of the RV \(S_i \triangleq Q_i-R_i\) is obtained as follows:
\vspace{-0.1cm}
\begin{equation}\label{eq: PDF of diff of chi}
    f_{S_i}(x)  \triangleq 0.25 \exp\left( -|x| /2\right).
\end{equation}
Finally, returning to \eqref{eq:expectation-proposition} and exploiting \eqref{eq:Chi-squared reform} and \eqref{eq: PDF of diff of chi}, we can derive the lower bound for expected achievable SNR as:
\begin{equation}\label{eq:expectation_final}
    \mathbb{E} \left[ |x_i z_i - y_i w_i| \right] = 0.25 \int_{-\infty}^{\infty} |x|f_S(x) \, dx = 0.5.
\end{equation}
The proof is concluded by substituting \eqref{eq:expectation_final} into \eqref{eq:expectation-proposition}.
\end{proof}
\end{theorem}


\begin{theorem}[Lower bound for the achievable SNR under pure LoS channels]\label{theorem:stochastic-LB-LoS}
Assume pure LoS channel conditions for all wireless links and let $\boldsymbol{h}_i = \exp{(\jmath \boldsymbol{\vartheta}^{\boldsymbol{h}}_i)}$
and $\boldsymbol{g}_i^* = \exp{(-\jmath \boldsymbol{\vartheta}^{\boldsymbol{g}}_i})$. The achievable deterministic SNR value $\gamma(\boldsymbol{\phi}^{\text{SA}})$ is again lower bounded by $\bar{f}_{\rm LB}(\boldsymbol{\phi}^{\text{SA}})$, similar to Theorem~\ref{theorem:stochastic-LB}.
\begin{proof}
Let $\boldsymbol{\vartheta}_i \triangleq \angle \boldsymbol{c}_i = \boldsymbol{\vartheta}^{\boldsymbol{h}}_i - \boldsymbol{\vartheta}^{\boldsymbol{g}}_i$ and assume, without loss of generality, that $\sum_{i=1}^{N} |\mathfrak{Re}(\boldsymbol{c}_i)|  > \sum_{i=1}^{N} |\mathfrak{Im}(\boldsymbol{c}_i)|$, which is equivalently expressed as $ \sum_{i=1}^{N} |\cos(\boldsymbol{\vartheta}_i)|  > \sum_{i=1}^{N} |\sin(\boldsymbol{\vartheta}_i)|$. The following derivations hold:
\vspace{-0.1 cm}
\begin{align}
    & \cos^2(\boldsymbol{\vartheta}_i) + \sin^2(\boldsymbol{\vartheta}_i)\!=\!1 \Rightarrow\! \sum_{i=1}^{N}\! \cos^2(\boldsymbol{\vartheta}_i) +  \sum_{i=1}^{N}\! \sin^2(\boldsymbol{\vartheta}_i) \!=\! N \nonumber \\
    & \Rightarrow\!
    \sum_{i=1}^{N}\! |\cos(\boldsymbol{\vartheta}_i)| +  \sum_{i=1}^{N}\! |\sin(\boldsymbol{\vartheta}_i)| \geq N 
    \Rightarrow\! \sum_{i=1}^{N}\! |\cos(\boldsymbol{\vartheta}_i)| \geq \frac{N}{2} \nonumber \\
    & \Rightarrow\! \left(\sum_{i=1}^{N}\! |\cos(\boldsymbol{\vartheta}_i)| \right)^{\!\!2} \geq 0.25N^2 \overset{\eqref{eq:lb1}} {\Rightarrow} \gamma(\boldsymbol{\phi}^{\text{SA}}) \geq 0.25N^2.
\end{align}
\vspace{-0.15 cm}
\end{proof}
\end{theorem}
\vspace{-0.35cm}
\begin{remark}
The lower bound $\bar{f}_{\rm LB}(\boldsymbol{\phi}^{\text{SA}})$ in Theorems~\ref{theorem:stochastic-LB} and~\ref{theorem:stochastic-LB-LoS} has an optimal asymptotic growth of $\Theta(N^2)$.
This is due to the fact that the continuous case of $\mathcal{OP}_1$, i.e., when $\boldsymbol{\theta}_n\in[0,2\pi]$ $\forall$$n$, can be optimally solved via Phase Alignment (PA)~\cite{PZZ22}, yielding the upper bound SNR value $N^2$ for normalized $\rho{\rm PL}$.
\end{remark}
\section{Iterative Discrete Optimization Schemes}\label{sec:lit-review}
As previously mentioned, for the perfect CSI availability case, the continuous version of $\mathcal{OP}_1$ can be optimally solved via PA \cite{PZZ22}, i.e., by setting $\boldsymbol{\phi}^{\rm PA}_n$ to $-(\angle \boldsymbol{h}_n - \angle \boldsymbol{g}_n)$ $\forall$$n$.
The most common approach in the literature for treating the discretized case is to optimally solve the continuous problem via PA, and then quantize each element so that $\boldsymbol{\phi}^{QPA}_n \triangleq {\rm arg}\min_{\boldsymbol{\phi}_n \in \mathcal{F}}\|\boldsymbol{\phi}_n - \boldsymbol{\phi}^{\rm PA}_n\|$~\cite{PZZ22, Risdiscrete2018, WZ20, ZWZ21}; this will be referred to as Quantized Phase Alignment (QPA) in the sequel.
It has been shown \cite{WZ20} that, under Rayleigh conditions, QPA achieves an asymptotic expected gain of about $0.4N^2$.

Other categories of discrete optimization algorithms include Branch and Bound (BnB) approaches that treat the quantization problem as mode selection \cite{HWC21, WZ20}. Despite their effectiveness, BnB methods have exponential computational requirements which makes them less attractive for real-time RIS configuration with instantaneous CSI. Black-box optimization algorithms, such as Genetic Optimization (GO), have also been used in discrete RIS tuning problems, such as in~\cite{PLP21, YCY23, SSH22}, with encouraging performance.
However, their population-based search entails many times more SNR evaluations than $N$ during a fixed CSI frame, which is infeasible for instantaneous RIS tuning. Those methods are in principal better tailored for statistical CSI setups~\cite{MAD23} or codebook search approaches~\cite{RIS_beam_management}.

From the perspective of deep learning, reinforcement learning algorithms for discrete \cite{ASH21} and $1$-bit \cite{SA22} RIS phase configuration have been lately proposed, as well as a two-step supervised learning methodology that couples the problems of SNR prediction and optimization~\cite{LHG23}.
Despite the fact that such approaches have demonstrated promising results in real-time RIS control, their performance is dependent on collected data that are scenario specific, and often imply considerable measurement overheads.
For this reason, these are not considered in the evaluation framework of this paper.


\subsection{Hill Climbing (HC)}\label{sec:hill-climbing}
One of the typical baseline approaches to solve the discrete optimization problem $\mathcal{OP}_2$ is to: \textit{i}) start with a random candidate solution; \textit{ii}) flip each RIS element's phase shift from $\phi^{0}$ to $\phi^{\pi}$ or vice-versa; \textit{iii}) evaluate the objective function (i.e., SNR) after the flip is performed; and \textit{v}) discard the change if no SNR improvement was obtained. This procedure is one of the standard generic discrete optimization approaches, commonly termed as ``coordinate hill climbing''~\cite{AI_Book}, which is described in Algorithm~\ref{alg:3} specifically for the problem of the $1$-bit RIS phase configuration selection. In this algorithm, we propose to use $\myvec{\phi}^{\text{SA}}$ from Algorithm~\ref{alg:1} as a starting RIS configuration. This has the two-fold benefit of starting the search from an already improved point, compared to random initialization, as well as stronger performance guarantees. Since the Hill Climbing (HC) algorithm may only improve on any given configuration, the performance bounds of Section~\ref{sec:sign-alignment-analysis} hold for Algorithm~\ref{alg:3}, as well.
We will be referring to the randomly initialized approach as HC, and as HC\&SA when initialized by  $\myvec{\phi}^{\text{SA}}$.
Notice that HC\&SA will outperform SA by design, albeit at an iterative computational cost.

\begin{algorithm}[t]
\caption{Iterative RIS Phase Configuration}\label{alg:3}
\begin{algorithmic}[1]
\Require The CSI vector $\myvec{c}$.
\State  Obtain $\myvec{\phi}^{(0)} \gets \myvec{\phi}^{\text{SA}}$ from Algorithm~\ref{alg:1}.
\State Set $\gamma^{\rm max} \gets \gamma(\myvec{\phi}^{(0)})$.
\For{$k=1,2,\ldots$ until convergence}
    \For{$n=1,2,\ldots,N$}
        \State Set $\myvec{\phi}^{(k)}_n \gets - \myvec{\phi}^{(k-1)}_n$.
    
        \If {$\gamma(\myvec{\phi}^{(k)}) > \gamma^{\rm max}$}
            \State Set $\gamma^{\rm max} \gets \gamma(\myvec{\phi}^{(k)})$.
        \Else
            \State Set $\myvec{\phi}^{(k)}_n \gets \myvec{\phi}^{(k-1)}_n$.
        \EndIf
    \EndFor
\EndFor
\State \Return $\myvec{\phi}^{(k)}$
\end{algorithmic}
\end{algorithm}
\section{Numerical Evaluation}
In this section, we evaluate the performance of the proposed SA RIS configuration methodology and compare it against: \textit{i}) QPA, as the main candidate closed-form solution available in the literature; \textit{ii}) the HC iterative algorithm of Section~\ref{sec:hill-climbing} that is comparably less computationally demanding than the relevant works discussed in Section~\ref{sec:lit-review}; and \textit{iii}) HC\&SA, to assess the improvements brought by the initialization via SA. We have also evaluated the SNR obtained via the upper bound approach of the continuous case through PA, which has been used for normalizing the achievable SNRs of all previous methods.
For the experimentation setup, we used the generic Ricean system model in~\cite{ASH21}, while the RIS was modeled as a $N^{\rm vert}\times N^{\rm hor}$ Uniform Rectangular Array (URA) so that $N = N^{\rm vert} N^{\rm hor}$ with $|N^{\rm vert} - N^{\rm hor}|$ selected to be minimized. In the Cartesian $(x,y,z)$ frame, the TX was positioned at $(0,0,6)$~m, the RX at $(0,20,1.5)$~m, and the RIS at $(2,2,2)$~m, with the latter's orientation being perpendicular to the $yz$ plane.
The performance curves included in this section are reported as averages over $1000$ random channel realizations, apart from Fig.~\ref{fig:snr_vs_N_LoS}, where realizations were averaged over $900$ deterministic channels of different angles, and from Fig.~\ref{fig:snr-vs-RIS-SMALL}, where $200$ realizations were averaged to alleviate from the consuming computations of exhaustive search.

\begin{figure}[t]
    \centering
    \includegraphics[width=0.7\linewidth]{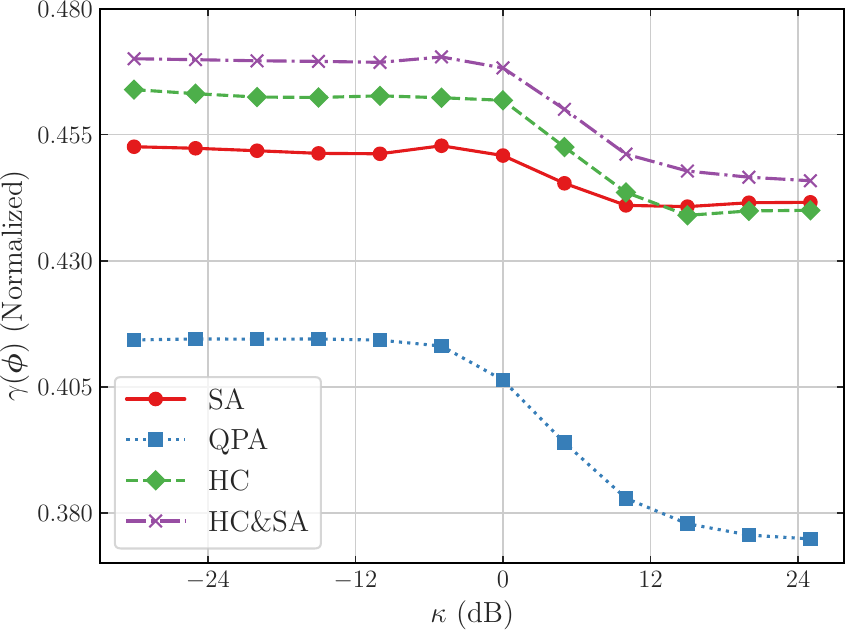}
    \caption{Average achievable SNR versus increasing Ricean $\kappa$-factors from Rayleigh fading to LoS conditions for an RIS with $N=100$ unit cells.\vspace{-0.2cm}}
    \label{fig:snr-vs-ricean-factor}
\end{figure}

In Fig.~\ref{fig:snr-vs-ricean-factor}, the performance comparison of the considered methods is reported under different Ricean $\kappa$-factors, namely, from $-30$dB (rich scattering) to $25$dB (almost pure LoS). As shown, SA achieves about $8$\% improvement on the normalized SNR over QPA under Rayleigh fading and around $17$\% improvements under LoS conditions. HC offers a mere $2$\% increase over SA under Rayleigh, while this difference is diminished for LoS, where SA performs equally well. HC\&SA offers a small improvement over HC, however, it comes at a faster convergence cost, as will be discussed in the sequel.

The results of Fig.~\ref{fig:snr-vs-ricean-factor} could be specific to the considered setup parameters, hence, we have performed averages over Ricean fading and all possible elevation/azimuth angles in Fig.~\ref{fig:snr_vs_N}, where the performances are evaluated as $N$ increases. Similar gains among all considered methods are observed in~Fig.~\ref{fig:snr_vs_N_rayleigh}, as in Fig.~\ref{fig:snr-vs-ricean-factor}, however, the mean SNR values among all methods are more closely comparable under LoS (Fig.~\ref{fig:snr_vs_N_LoS}), when all possible angles are averaged out, indicating that the problem of selecting effective RIS configuration schemes is tightly coupled with its positioning under LoS conditions. It is important to highlight that in~Fig.~\ref{fig:snr_vs_N_rayleigh}, QPA converges to its asymptotic normalized expected gain of $0.4$ for larger $N$, as stated in Section~\ref{sec:lit-review}, however, SA immensely exceeds the lower bounds given in Section~\ref{sec:sign-alignment-analysis}. The convergence iterations for Figs.~\ref{fig:snr_vs_N_rayleigh} and~\ref{fig:snr_vs_N_LoS} for the two proposed iterative approaches are given in Table~\ref{tab:convergence-iters}, from which it can be inferred that, in RISs with $N\geq100$, the SA initialization offers $15$\%-$25$\% convergence speed improvements, on top of marginally improved performance.

The performance evaluation displayed so far has been compared against the SNR achievable by an RIS with continuous phase shifts, which calls for investigation on the performance of the considered methods with respect to the optimal $1$-bit-quantized solution of $\mathcal{OP}_1$.
For $N$ up to $26$, where evaluating all $2^N$ combinations is computationally tractable, this comparison is given in Fig.~\ref{fig:snr-vs-RIS-SMALL} for Rayleigh fading conditions.
Interestingly, in the small RIS regime, SA obtains around $95$\% of the optimal performance, despite being a closed-form approximate solution, which hints that the results in larger RIS sizes may not be too far from the optimal value. Evidently, the proposed iterative approaches obtain near-optimal performance in these scenarios, however, there is a maintained $13$\%-$15$\% performance difference between SA and QPA, indicating that the proposed closed-form RIS configuration is offering firm benefits with negligent computational overhead.
\begin{figure*}
\centering
\subfloat[Rayleigh fading]{\includegraphics[width=0.29\linewidth]{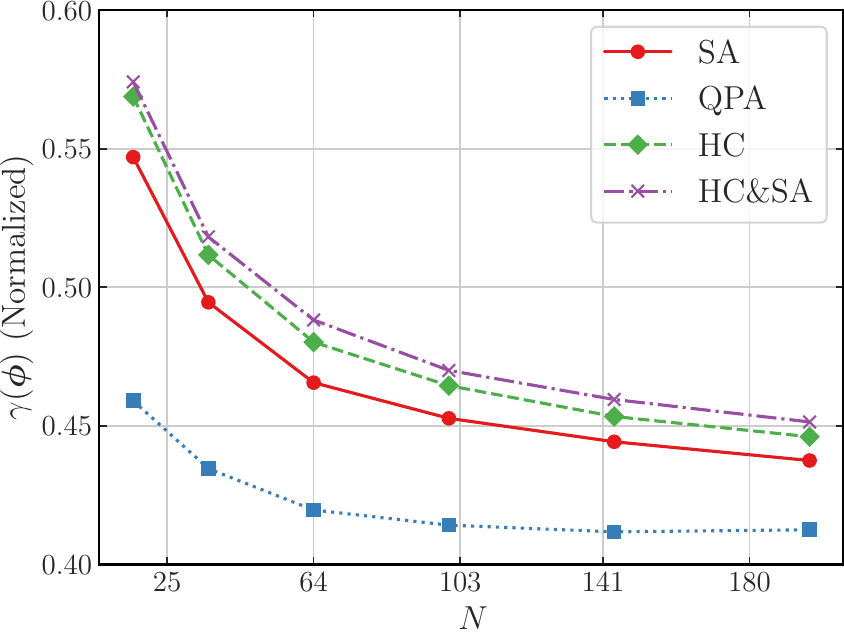}\label{fig:snr_vs_N_rayleigh}}~
\subfloat[Pure LoS conditions]{\includegraphics[width=0.29\linewidth]{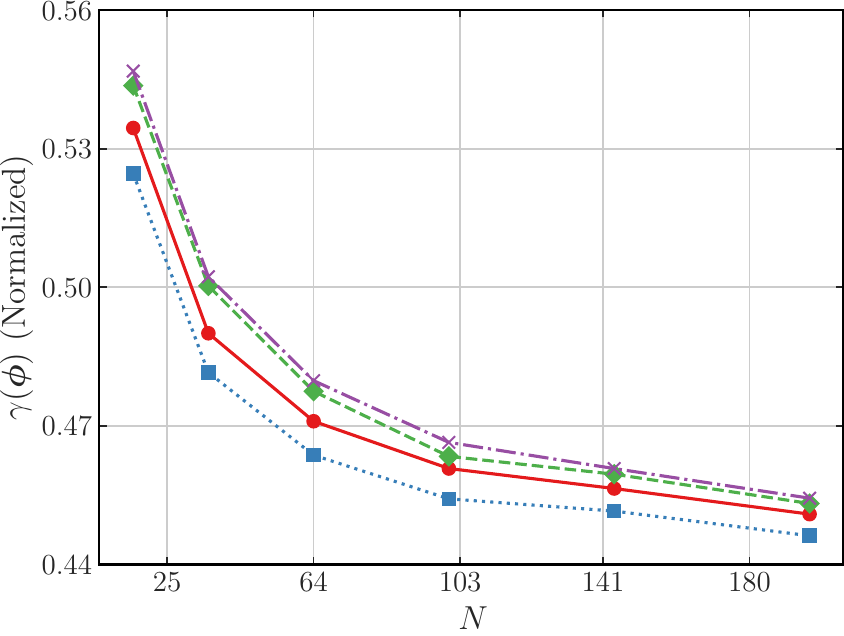}\label{fig:snr_vs_N_LoS}}~
\subfloat[Rayleigh fading (few-element RIS)]
{\includegraphics[width=0.29\linewidth]{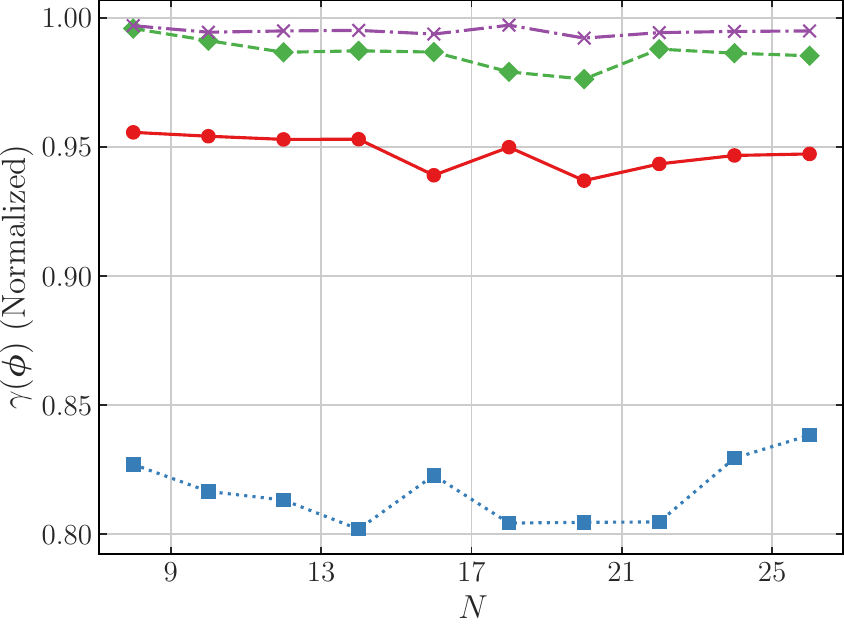}\label{fig:snr-vs-RIS-SMALL}}
\caption{Average achievable SNR versus the number of RIS unit cells under different fading conditions. The results are normalized, similar to Fig$.$~\ref{fig:snr-vs-ricean-factor}, w.r.t.: \textit{i}) the upper bound SNR obtainable by non-quantized phase shifts in (a) and (b); and \textit{ii}) the SNR obtained by exhaustively solving $\mathcal{OP}_1$ in (c).\vspace{-0.45cm}}
\label{fig:snr_vs_N}
\end{figure*}

\begin{table}[t]
    \centering
    \caption{Average Number of HC Iterations for Convergence.}
    \begin{tabular}{c|*{6}{c}}
        \hline
        $N$ & {$16$} & {$36$} & {$64$} & {$100$} & {$144$} & {$196$} \\
        \hline
        &\multicolumn{6}{c}{Rayleigh} \\
        HC & $47.52$ & $120.78$ & $254.72$ & $421.5$ & $648.0$ & $1030.96$ \\
        HC\&SA & $47.68$ & $113.22$ & $209.6$ & $355.5$ & $551.52$ & $814.38$ \\
        &\multicolumn{6}{c}{LoS} \\
        HC  & $48.79$ & $126.0$ & $232.49$ & $372.53$ & $635.11$ & $851.15$ \\
        HC\&SA & $40.44$ & $103.44$ & $191.80$ & $267.59$ & $478.22$ & $715.04$ \\
        \hline
    \end{tabular}
    \label{tab:convergence-iters}
\end{table}

\section{Implementation of RIS Sign Alignment}
Notwithstanding its performance improvements over conventional RIS configurations, SA can offer benefits over the related QPA closed-form approach also in terms of hardware requirements. To compute $\boldsymbol{\phi}^{\rm Re}$ and $\boldsymbol{\phi}^{\rm Im}$ from Algorithm~\ref{alg:1}, one needs only the signs of each channel gain $\boldsymbol{c}_i$, which is equivalent to measuring the quadrant of $\angle \boldsymbol{c}_i$. For each single-hop channel, those signs can be acquired through a simple Reception (RX) Radio-Frequency chain (RFC) comprising a pair of $1$-bit Analog-to-Digital Converters (ADCs), tasked to measure the signs of each $\boldsymbol{c}_i$'s I/Q components \cite{DMA_1bit}. Note that the power consumption of ADCs grows exponentially with their bit resolution, hence, an $1$-bit ADC offers an important improvement over traditional high-resolution ADCs. Secondly, since no amplitude information is required to compute the SA configurations, the detection process can be further simplified and will be less susceptible to errors, equivalently to how Quadrature-Phase-Shift-Keying (QPSK) constellations are easier to detect compared to high-order Quadrature Amplitude Modulation (QAM)~\cite{goldsmith2005wireless}.

On a conceptual level, such simplified and low-power RX RFCs can be envisioned to be endowed onto the metasurface, toward autonomous RIS solutions~\cite{alexandropoulos2023hybrid}. In fact, the two output bits of the two ADCs, could be directly connected to the RIS element switches, so that 
the surface's state may be set to $\boldsymbol{\phi}^{\rm Re}$ or $\boldsymbol{\phi}^{\rm Im}$ without the need for involved digital controllers. Since each element's configuration is independent of the other channel coefficients, the operation can take place at the unit-cell level. Furthermore, the design is straightforwardly related to the output of the ADCs, thus no detection algorithm is needed at the RIS controller~\cite{ABoI_EURASIP}. Notice that the SNR computations implied by Step~$3$ of Algorithm~\ref{alg:1} could be performed at the receiver, where high-resolution RX RFCs can be readily available, and the result of the comparison could be fed back to the RIS if a control channel is present. However, describing a detailed RX RFC and its integration to the RIS, that can facilitate SA on the RIS side with reduced hardware complexity, lies beyond the scope of this paper.


It is finally noted that, from a communications system perspective, the considered SISO system may be extended to the general case of Multiple Input Multiple Output (MIMO) systems by leveraging a capacity analysis on the eigenvalues and eigenvectors of the covariance matrices of the wireless channels, as it has been recently studied in~\cite{MAD23}.
We delegate such analysis to the journal version of this work.


\section{Conclusion}

In this paper,  we considered an RIS-enabled SISO communication system and presented a novel closed-form configuration for the RIS responses for the case of $1$-bit-phase-quantized unit cells where the phase difference is of $\pi$ radians. Lower bounds of the proposed SA configuration under Rayleigh fading and LoS conditions, which are asymptotically optimal, were derived. Our numerical evaluation showcased that SA outperforms the state-of-the-art configuration scheme, which is obtained by quantizing the optimal continuous phase shifts. It was also demonstrated that when SA is used as initial point in the iterative HC algorithm, it can provide further performance increase with faster convergence. We finally discussed how the proposed closed-form RIS configuration can be realized at the RIS side in a standalone operation fashion via low-power RX units each comprising a pair of $1$-bit ADCs.

\vspace{-0.1cm}

\FloatBarrier
\bibliographystyle{IEEEtran}
\vspace{-0.1cm}
\bibliography{references}
\vspace{-0.05cm}

\end{document}